\documentclass{emulateapj}
\usepackage{color}

\lefthead{Faltenbacher et al.}
\righthead{Galaxy alignment within dark matter halos}

\newcommand{\hMsol}{{\,h^{-1}\rm M}_\odot}

\newcommand{\hkpc}{{\,h^{-1}\rm kpc}}
\newcommand{\kms}{{\,\rm km~s^{-1}}}
\newcommand{\Rvir}{{\,R_{\rm vir}}}
\newcommand{\Om}{\Omega_{\rm m}}
\newcommand{\Ol}{\Omega_\Lambda}

\begin{document}
\title{Three Different  Types of  Galaxy Alignment within  Dark Matter
Halos}
\author {A. Faltenbacher\altaffilmark{1}, Cheng Li\altaffilmark{1}, 
Shude Mao\altaffilmark{2}, Frank C. van den Bosch\altaffilmark{3}, 
Xiaohu Yang\altaffilmark{1}, Y.P. Jing\altaffilmark{1},
Anna Pasquali\altaffilmark{3} and H.J. Mo\altaffilmark{4}}
\altaffiltext{1} {Shanghai  Astronomical Observatory,  Nandan Road 80,
Shanghai 200030, China}
\altaffiltext{2} {University of Manchester, Jodrell Bank Observatory,
Macclesfield, Cheshire SK11 9DL, UK}    
\altaffiltext{3} {Max-Planck-Institute for Astronomy, K\"onigstuhl 17,
D-69117 Heidelberg, Germany }
\altaffiltext{4}{Department of Astronomy, University of Massachusetts,
Amherst MA 01003-9305}
\begin{abstract}
  Using a large galaxy group catalogue  based on the Sloan Digital Sky
  Survey Data Release 4 we measure  three different types of intrinsic
  galaxy  alignment  within   groups:   halo   alignment  between  the
  orientation    of  the  brightest   group  galaxies   (BGG)  and the
  distribution of its satellite galaxies, radial alignment between the
  orientation of a satellite galaxy and the direction towards its BGG,
  and direct alignment between the orientation  of the BGG and that of
  its   satellites.  In agreement  with  previous studies we find that
  satellite galaxies are preferentially  located along the major axis.
  In addition, on scales $r  < 0.7 \Rvir$ we  find that red satellites
  are  preferentially aligned radially  with the direction to the BGG.
  The   orientations   of  blue   satellites,  however,  are perfectly
  consistent with being isotropic. Finally, on scales $r < 0.1 \Rvir$,
  we find a  weak but   significant  indication for  direct  alignment
  between  satellites and BGGs.   We  briefly discuss the implications
  for weak lensing measurements.
\end{abstract}
\section{Introduction}
A  precise assessment of  galaxy alignments is  important for two main
reasons: it  contains information regarding  the impact of environment
on the formation and evolution of galaxies, and it can be an important
source of contamination for weak lensing measurements.  In theory, the
large scale-tidal field is expected to induce large-scale correlations
between          galaxy        spins   and        galaxy        shapes
\citep[e.g.,][]{2000ApJ...543L.107P,              2000ApJ...545..561C,
  2000MNRAS.319..649H,     2001MNRAS.320L...7C,   2001ApJ...559..552C,
  2002MNRAS.332..339P,  2002MNRAS.335L..89J}.      In    addition, the
preferred  accretion of  new material along   filaments tends to cause
alignment with  the  large scale filamentary  structure  in which dark
matter      halos        and        galaxies       are        embedded
\citep[e.g.,][]{2002MNRAS.335L..89J,              2005MNRAS.362.1099F,
  2005ApJ...627..647B}.  On small  scales,  however, inside virialized
dark matter haloes, any  primordial alignment is  likely to have  been
significantly weakened due to   non-linear  effects such  as   violent
relaxation         and            (impulsive)               encounters
\citep[e.g.,][]{2002MNRAS.332..325P}.  On the other hand, tidal forces
from the host halo   may also induce  new  alignments, similar to  the
tidal    locking  mechanism  that    affects   the  Earth-Moon  system
\citep[e.g.,][]{1994MNRAS.270..390C,              1997ApJ...487..489U,
  2003ApJ...592..147F}.

Observationally, the search for galaxy alignments has a rich and often
confusing history.  To some extent this owes to the fact that numerous
different forms  of alignment have  been discussed in  the literature:
the        alignment        between       neighbouring        clusters
\citep{1982A&A...107..338B, 1989ApJ...344..535W, 1994ApJS...95..401P},
between brightest  cluster galaxies  (BCGs) and their  parent clusters
\citep{1980MNRAS.191..325C, 1982A&A...107..338B, 1990AJ.....99..743S},
between the  orientation of satellite galaxies and  the orientation of
the  cluster   \citep{1985ApJ...298..461D,  2003ApJ...594..144P},  and
between the  orientation of satellite galaxies and  the orientation of
the  BCG  \citep{1990AJ.....99..743S}.   Obviously, several  of  these
alignments   are   correlated  with   each   other,  but   independent
measurements are  difficult to compare  since they are often  based on
very different data sets.

With  large galaxy  redshift  surveys, such  as  the two-degree  Field
Galaxy Redshift  Survey \citep[2dFGRS,][]{2001MNRAS.328.1039C} and the
Sloan Digital Sky  Survey \citep[SDSS,][]{2000AJ....120.1579Y}, it has
become possible to investigate  alignments using large and homogeneous
samples. This has resulted in robust detections of various alignments:
\cite{2005ApJ...628L.101B},       \cite{2006MNRAS.369.1293Y}       and
\cite{2007MNRAS.376L..43A}  all  found  that  satellite  galaxies  are
preferentially  distributed  along  the   major  axes  of  their  host
galaxies,   \cite{2006ApJ...640L.111T}  found  that   spiral  galaxies
located  on the  shells of  large voids  have rotation  axes  that lie
preferentially on the void surface, and \cite{2005ApJ...627L..21P} and
\cite{2006ApJ...644L..25A} noticed that  satellite galaxies tend to be
preferentially oriented towards the galaxy at the center of the halo.

In this Letter we use  a large galaxy group catalogue constructed from
the SDSS to study galaxy alignments on small scales within dark matter
haloes that  span a wide range  in masses.  The unique  aspect of this
study is that we investigate  three different types of alignment using
exactly the  same data  set consisting of  over $60000$  galaxies.  In
addition, by using a carefully selected galaxy group catalogue, we can
discriminate between central galaxies  and satellites, and study their
mutual   alignment.   The   latter  is   particularly   important  for
galaxy-galaxy  lensing,  where  it  can  be a  significant  source  of
contamination.  Finally,  exploiting the  large number of  galaxies in
our sample,  we also investigate  how the alignment signal  depends on
the colors of the galaxies. Throughout we adopt $\Om = 0.3$ and $\Ol =
0.7$ and a Hubble parameter $h = H_0/100\kms{\rm Mpc}^{-1}$.
\section{Data \& Methodology}
\label{sec:data}
\begin{figure}
\plotone{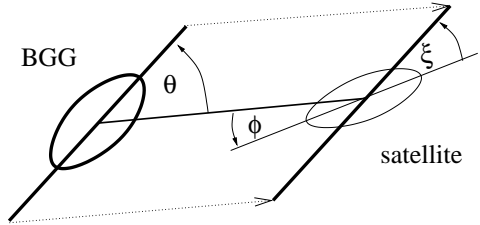}
\caption{\label{fig:sketch}
  Illustration of  the three angles $\theta$, $\phi$  and $\xi$, which
  are used  to test  for halo alignment,  radial alignment  and direct
  alignment, respectively. The three angles are not independent: if
  ordered by size $\alpha\geq\beta\geq\gamma$ then 
  $\alpha=\min[\beta+\gamma,180^\circ-\beta-\gamma]$.}
\end{figure}
We apply our  analysis to the SDSS  galaxy group catalogue  of Yang et
al.   (2007,  in prep.).   This   catalogue is  constructed  using the
halo-based group finder  of \cite{2005MNRAS.356.1293Y} and  applied to
the New York University Value Added Galaxy Catalog (NYU-VAGC)
\footnote{http://wassup.physics.nyu.edu/vagc/}  that is  based  on the
SDSS  Data   Release   Four \citep[DR4;][]{2006ApJS..162...38A}.  This
group  finder uses the general properties  of  CDM halos (i.e.  virial
radius, velocity dispersion, etc.)  to  determine the memberships   of
groups \citep[cf.][]{2006MNRAS.366....2W}.  In this study we only
use those groups with redshifts in  the range $0.01\leq z\leq 0.2$ and
with   halo  masses between    $5\times  10^{12}\hMsol$ and $5  \times
10^{14}\hMsol$.   In addition,  we  only focus  on group  members with
$^{0.1}M_r - 5\log h \leq  -19$.  Throughout this paper all magnitudes
are $k+e$ corrected  to $z=0.1$ following  \cite{2003ApJ...592..819B}.
Using the method of
\cite{2006MNRAS.368...21L} we split our galaxies  in three color bins.
In short,  we   divide the  full NYU-VAGC   sample  in 282  subsamples
according to the $r$-band luminosity,  and fit the $^{0.1}(g-r)$ color
distribution for  each subsample with  a double-Gaussian.  Galaxies in
between the centers  of the two  Gaussians are classified as  `green',
while those with higher and  lower values for the $^{0.1}(g-r)$  color
are  classified as `red'  and `blue', respectively.  The final sample,
on which our  analysis is  based, consists  of $18576$  groups with  a
total of $60724$  galaxies,   of which  $29780$ are red,   $20604$ are
green, and $10340$ are blue.

In what    follows, we use    these groups to   examine  (i) {\it halo
alignment} between  the orientation  of  the brightest  group galaxies
(BGG) and the distribution of its satellite galaxies, (ii) {\it radial
alignment} between the  orientation   of a  satellite galaxy and   the
direction towards  its BGG, and  (iii) {\it direct  alignment} between
the orientation of the BGG and that of its satellites.  In particular,
we define  the angles  $\theta$,  $\phi$ and  $\xi$ as  illustrated in
Fig.~\ref{fig:sketch}, and investigate whether their distributions are
consistent   with isotropy,  or whether   they   indicate a  preferred
alignment.  Following \cite{2005ApJ...628L.101B} and
\cite{2006MNRAS.369.1293Y}, the orientation of  each galaxy is defined
by the  major  axis position angle  (PA) of  its 25-magn arcsec$^{-2}$
isophote in the $r$-band.

For each satellite  galaxy we compute its projected  distance, $r$, to
the BGG,  normalized by the virial  radius, $\Rvir$, of  its group (as
derived  from the group  mass).  For  each of  5 radial  bins, equally
spaced  in  $r/\Rvir$,  we  then  compute  $\langle  \theta  \rangle$,
$\langle \phi  \rangle$ and  $\langle \xi \rangle$,  where $\langle  . 
\rangle$ indicates the average over all BGG-satellite pairs in a given
radial  bin.   Next we  construct  100  random  samples in  which  the
positions  of  the  galaxies  are   kept  fixed,  but  their  PAs  are
randomized.   For each  of these  random samples  we  compute $\langle
\theta \rangle$,  $\langle \phi \rangle$ and $\langle  \xi \rangle$ as
function of $r/\Rvir$, which we use to compute the significance of any
detected alignment signal.
\section{Results}
\label{sec:results}
\subsection{Halo alignment}
\label{sec:halo}
\begin{figure}
\plotone{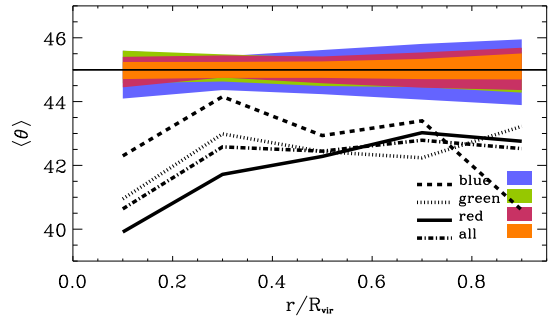}
\caption{\label{fig:Theta01_4}
  Mean  angle, $\theta$,  between  the  PA  of  the BGG and   the line
  connecting  the BGG   with   a satellite  galaxy,  as  function   of
  $r/\Rvir$.   Different line  styles indicate (sub)samples determined
  according  to  the  satellites'  color.  The shaded  areas  mark the
  parameter  space  between the $16^{\rm     th}$  and $84^{\rm   th}$
  percentiles  of the   distributions  obtained  from the   100 random
  samples.   A signal outside  this shaded   region means  that  it is
  inconsistent with no alignment (i.e., with isotropy) at more than 68
  percent confidence.}
\end{figure}
Fig.~\ref{fig:Theta01_4} shows the results thus obtained for the angle
$\theta$ between the  orientation of the BGG  and  the line connecting
the BGG with the satellite galaxy. Clearly, for all four samples shown
(all, red, green  and  blue, where the  color  refers to that   of the
satellite  galaxy, not that  of  the  BGG)  we obtain  $\langle \theta
\rangle  <  45^{\circ}$    at   all  5   radial   bins   and   at high
significance\footnote{More than 99 percent,  except for the $0.3\Rvir$
  bin for the blue and the $0.9\Rvir$ bin  for the green satellites.}.
This indicates  that satellite galaxies are preferentially distributed
along the major  axis of the BGG, in  good agreement with the findings
of    \cite{2005ApJ...628L.101B},   \cite{2006MNRAS.369.1293Y}     and
\cite{2007MNRAS.376L..43A},    but    opposite         to  the     old
\cite{1969ArA.....5..305H} effect.   Note  that    there is  a   clear
indication that the distribution of  red  satellites is more  strongly
aligned with the orientation of the BGG than  that of blue satellites,
again      in      good      agreement   with       previous   studies
\citep[cf.][]{2006MNRAS.369.1293Y, 2007MNRAS.376L..43A}
\subsection{Radial alignment}
\label{sec:radial}
\cite{1975AJ.....80..477H}  were  the   first  to  report  a  possible
detection  of  radial  alignment   in  the  Coma  cluster,  which  has
subsequently   been   confirmed   by  \cite{1976ApJ...209...22T}   and
\cite{1983ApJ...274L...7D}.  However, in a more systematic study based
on the  2dFGRS, \cite{2002AJ....124..733B}  were unable to  detect any
significant  radial alignment  of satellite  galaxies  around isolated
host galaxies.  On  the other hand, using a  very similar selection of
hosts     and    satellites,    but     applied    to     the    SDSS,
\cite{2006ApJ...644L..25A}  found   significant  evidence  for  radial
alignment    on    scales    $\lesssim   70\hkpc$.     In    addition,
\cite{2005ApJ...627L..21P}  found   a  statistically  robust  tendency
toward  radial  alignment in  a  large  sample  of 85  X-ray  selected
clusters.

\begin{figure}
\plotone{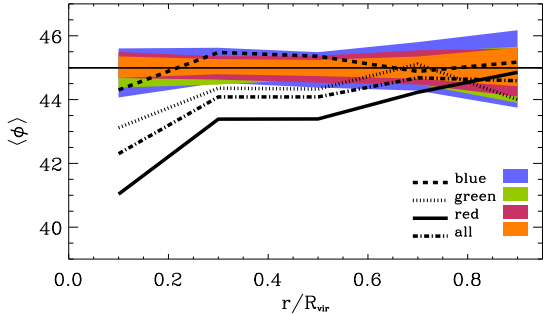}
\caption{\label{fig:Phi01_4}
  Same  as Fig.~\ref{fig:Theta01_4},  but  for the  angle $\phi$  (see
  Fig.~\ref{fig:sketch}).}
\end{figure}
Fig.~\ref{fig:Phi01_4}  shows  the  results  obtained from  our  group
catalogue.  It shows, as function  of $r/\Rvir,$ the mean angle $\phi$
between the PA of the  satellite and the line connecting the satellite
with its BGG. As in Fig.~\ref{fig:Theta01_4} results are shown for all
four different  samples, together with the $16^{\rm  th}$ and $84^{\rm
th}$ percentiles obtained from the random samples.  There is a clear
and very significant indication that  the major axes of red satellites
point  towards the  BGG  (i.e., $\langle\phi\rangle  < 45^\circ$),  at
least for  projected radii $r  \lesssim 0.7\Rvir$. The signal  for the
green  satellites  is  significantly   weaker,  but  still  reveals  a
preference for  radial alignment on small  scales: in fact,  for the 3
radial bins  with $r \leq  0.5\Rvir$ the null-hypothesis of  no radial
alignment can  be rejected at more  than 95 percent  confidence level. 
In contrast,  for the blue  galaxies the data is  perfectly consistent
with no radial alignment. Since the 2dFGRS is more biased towards blue
galaxies  than the  SDSS,  this  may at  least  partially explain  why
\cite{2002AJ....124..733B}  were unable  to detect  significant radial
alignment.

\subsection{Direct alignment}
\label{sec:direct}
The search for direct  alignment has mainly  been restricted to galaxy
clusters   \citep[e.g.,][]{2003ApJ...594..144P,   2005MNRAS.359..191S,
  2007astro.ph..3443T},  mostly     resulting  in  no  or    very weak
indications  for alignment   between  the   orientations of   BCG  and
satellite  galaxies.  \cite{2006ApJ...644L..25A} extended the   search
for direct  alignment   to a samples  of  4289  host-satellites  pairs
selected  from the SDSS DR4, finding  a weak but significant signal on
scales $\lesssim 35\hkpc$. On  larger scales, however,  no significant
alignment was found, in agreement with \cite{2006MNRAS.367..611M}.

\begin{figure}
\plotone{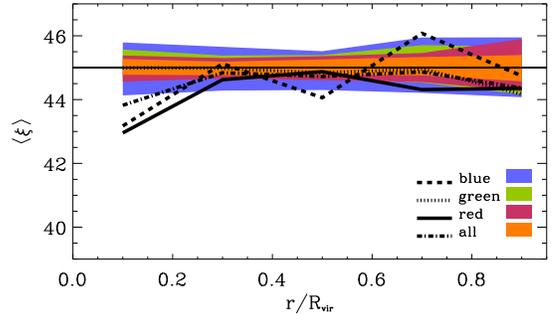}
\caption{\label{fig:Xi01_4}
  Same  as Fig.~\ref{fig:Theta01_4},  but  for the  angle $\xi$  (see
  Fig.~\ref{fig:sketch}).}
\end{figure}

Fig.~\ref{fig:Xi01_4} displays  our results  for the direct alignment,
based  on the angle  $\xi$ between   the orientations  of  a satellite
galaxy and  that of its BGG.   With  the exception of  the central bin
($r/\Rvir = 0.1$) the null-hypothesis of a random distribution cannot
be rejected at more than $1 \sigma$ confidence level. Our study, based
on over    40000   BGG-satellite  pairs,    therefore   agrees    with
\cite{2006ApJ...644L..25A} that there is a  weak indication for direct
alignment, but only on relatively small scales:  for the average group
mass in  our   sample,  $M  =  3.6\times10^{13}\hMsol$,  a   radius of
$r=0.1\Rvir$  corresponds to $70\hkpc$.  However, at least for the red
satellites there is   a  systematic trend towards   angles $<45^\circ$
which      may     be    caused   by      the     group   tidal  field
\citep[cf.][]{2005ApJ...629L...5L}.
\subsection{Dependence on selection criteria}
The sample used above is based  on galaxies with $^{0.1}M_r - 5\log h
\leq -19$.  Typically, including  fainter galaxies improves the number
statistics but  not necessarily the  signal-to-noise since the  PAs of
fainter galaxies carry  larger errors. To test the  sensitivity of our
results,  we repeated  the above  analysis using  magnitude  limits of
$-17$, $-18$, and $-20$.  This resulted in alignment signals that were
only marginally different. We have  also tested the sentitivity of our
results to  the range of  group masses considered. Changing  the lower
limit  to $10^{12} \hMsol$  or $10^{13}\hMsol$,  or imposing  no upper
mass limit,  all yields very  similar alignment signals.   These tests
assure that our selection criteria lead to representative results.

\section{Discussion}
\label{sec:diss}
The origin  of the halo  alignment described in  \S~\ref{sec:halo} has
been       studied       by       \cite{2006ApJ...650..550A}       and
\cite{2007astro.ph..1130K}  using  semi-analytical  models  of  galaxy
formation combined with large $N$-body simulations.  Since dark matter
haloes  are  in  general  flattened,  and  satellite  galaxies  are  a
reasonably fair tracer of  the dark matter mass distribution, $\langle
\theta \rangle$ will  be smaller than $45^{\circ}$ as  long as the BGG
is   aligned   with   its    dark   matter   halo.    In   particular,
\cite{2007astro.ph..1130K} were able  to accurately reproduce the data
of \cite{2006MNRAS.369.1293Y} under the assumption that the minor axis
of the BGG is perfectly aligned  with the spin axis of its dark matter
halo.

\cite{2007astro.ph..1130K}  also showed that  the color  dependence of
the  halo alignment  has a  natural  explanation in  the framework  of
hierarchical  structure   formation:  red  satellites   are  typically
associated  with subhaloes  that were  more massive  at their  time of
accretion.  Since  the orientation  of a halo  is correlated  with the
direction   along    which   it   accreted   most    of   its   matter
\citep[e.g.,][]{2005MNRAS.364..424W,     2005MNRAS.363..146L},     red
satellites are  a more  accurate tracer of  the halo  orientation than
blue satellites.

The origin of  the radial alignment is less  clear. One possibility is
that it reflects a left-over from large-scale alignments introduced by
the large  scale tidal  field and  the  preferred accretion of  matter
along filaments.  Such alignment,  however, is unlikely to survive for
more  than  a few orbits   within  the halo  of  the  BGG, so that the
observed  alignment must be mainly due  to the satellite galaxies that
were accreted most recently.   Since these satellites typically reside
at  relatively  large  halo-centric  radii,  this picture  predicts  a
stronger radial alignment at larger radii, clearly opposite to what we
find.

A   more likely explanation,  therefore, is  that radial alignment has
been    created  locally by  the  group   tidal  field.   As shown  by
\cite{1994MNRAS.270..390C}, the timescale on which a  prolate galaxy
can adjust its  orientation to  the tidal  field of a  cluster is much
shorter than the Hubble time, but  longer than its intrinsic dynamical
time.   Consequently, prolate  galaxies  have  a  tendency  to  orient
themselves towards the  cluster  center.  The fact  that  the observed
signal      increases  towards   the     group  center supports   this
interpretation.  In particular,  satellites  that were accreted  early
not  only  are more likely  to  be red,  they  also are more likely to
reside   at small  group-centric radii    and to  have relatively  low
group-centric  velocities \citep[e.g.,][]{2004ApJ...616..745M}.   This
will enhance their tendency to align themselves  along the gradient in
the cluster's gravitational potential, and  they may well be the major
contributors to the pronounced signal on small scales.  In the case of
disk galaxies, the conservation of intrinsic angular momentum prevents
the disk from  re-adjusting to the tidal  field, which may explain why
blue satellites  show no sign of radial  alignment. Finally, the tidal
field  of the  parent  halo also results in   tidal stripping, and the
tidal debris may influence  the inferred orientation of the  satellite
galaxy         \citep[cf.][]{2001ApJ...557..137J,2006MNRAS.366.1012F}.
Detailed studies  are  required to investigate  the  interplay between
intrinsic satellite orientations and the groups tidal field.

In order to understand the   direct alignment results, first   realize
that the angles $\theta$,  $\phi$ and $\xi$   are not independent  (see 
Fig.~\ref{fig:sketch}). However,  the equation given  in the caption is
only applicable for single cases not for  the mean angles. Our results
indicate  that  satellite  galaxies   are more likely  to  be  aligned
`radially'  with the direction towards the  BGG,  than `directly' with
the orientation  of the  BGG.   Since there  is no  clear  theoretical
prediction for direct alignment, at least not one that can survive for
several orbital periods in a dark matter  halo, while radial alignment
can  be  understood as  originating from   the halo's  tidal field, we
consider the relative weakness of  direct  alignment to be  consistent
with expectations.

In recent years  galaxy-galaxy (GG) lensing  has emerged  as a primary
tool for constraining the masses of  dark matter halos around galaxies
\citep[e.g.,][]{2004AIPC..743..129B}.      If satellite galaxies   are
falsely identified  as sources lensed by the  BGG, which  is likely to
happen in  the absence of redshift  information, the  radial alignment
detected here will dilute the  tangential GG lensing signal induced by
the  dark matter  halo associated with  the  BGG, thus resulting in an
underestimate   of    the   halo     mass.     In    agreement    with
\cite{2006ApJ...644L..25A}, our   findings  therefore emphasize    the
importance of an accurate rejection  of satellite galaxies to  achieve
precision constraints on    dark matter halo  masses from   GG lensing
measurements.  Similarly, the weak but significant detection of direct
alignment may contaminate the cosmic shear measurements. Since we only
detected a weak  signal  on small  scales, one can  easily  avoid this
contamination  by  simply  removing or  down-weighting close  pairs of
galaxies in projection \citep{2002A&A...396..411K, 2003MNRAS.339..711H}.
\vspace{-0.02\vsize}
\section*{Acknowledgments}
This work is supported  by NSFC (10533030, 0742961001, 0742951001) and
the Knowledge Innovation Program of  the Chinese Academy of  Sciences,
grant KJCX2-YW-T05. AF  and CL are supported  by the Joint Program
in  Astrophysical  Cosmology    of   the Max   Planck   Institute  for
Astrophysics  and the   Shanghai  Astrophysical Observatory.   YPJ  is
partially  supported by  Shanghai  Key   Projects in  Basic   research
(04JC14079 and 05XD14019).

\end{document}